# LCLS-II-HE verification cryomodule high gradient performance and quench behavior

S. Posen[1a], A. Cravatta[1], M. Checchin[1], S. Aderhold[2], C. Adolphsen[2], T. Arkan[1], D. Bafia[1], A. Benwell[2], D. Bice[1], B. Chase[1], C. Contreras-Martinez[1], L. Dootlittle[3], J. Fuerst[2], D. Gonnella[2], A. Grassellino[1], C. Grimm[1], B. Hansen[1], E. Harms[1], B. Hartsell[1], G. Hays[2], J. Holzbauer[1], S. Hoobler[2], J. Kaluzny[1], T. Khabiboulline[1], M. Kucera[1], D. Lambert[1], R. Legg[2], F. Lewis[1], J. Makara[1], H. Maniar[1], J. T. Maniscalco[2], M. Martinello[1], J. Nelson[2], S. Paiagua[3], Y. Pischalnikov[1], P. Prieto[1], J. Reid[1], M. Ross[2], C. Serrano[3], N. Solyak[1], A. Syed[1], D. Sun[1], G. Tatkowski[1], R. Wang[1], M. White[1], L. Zacarias[2]

[1]*Fermi National Accelerator Laboratory, Batavia, Illinois, 60510, USA*
[2]*SLAC National Accelerator Laboratory, Menlo Park, California 94025, USA*
[3]*Lawrence Berkeley National Laboratory, Berkeley, California 94720, USA*

An 8-cavity, 1.3 GHz, LCLS-II-HE cryomodule was assembled and tested at Fermilab to verify performance before the start of production. Its cavities were processed with a novel nitrogen doping treatment to improve gradient performance. The cryomodule was tested with a modified protocol to process sporadic quenches, which were observed in LCLS-II production cryomodules and are attributed to multipacting. Dedicated vertical test experiments support the attribution to multipacting. The verification cryomodule achieved an acceleration voltage of 200 MV in continuous wave mode, corresponding to an average accelerating gradient of 24.1 MV/m, significantly exceeding the specification of 173 MV. The average $Q_0$ ($3.0 \times 10^{10}$) also exceeded its specification ($2.7 \times 10^{10}$). After processing, no field emission was observed up to the maximum gradient of each cavity. This paper reviews the cryomodule performance and discusses operational issues and mitigations implemented during the several month program.

## I. INTRODUCTION

The LCLS-II-HE [1] project will increase the beam energy of the LCLS-II (Linac Coherent Light Source-II) [2] superconducting linac from 4 GeV to 8 GeV. More than 20 LCLS-II-HE cryomodules will be built by Fermilab and Jefferson Lab, each containing eight 1.3 GHz, 9-cell, TESLA-style [3] superconducting radiofrequency (SRF) cavities. The cryomodule design (see Figure 1) is basically unchanged from that for the LCLS-II 'L2' modules. However, there are significant changes in the cavity performance requirements (see Table 1). The required average accelerating gradient was increased from 16.0 MV/m in L2 to 20.8 MV/m in HE, corresponding to a total module voltage of 173 MV. This increase was motivated by tunnel length constraints and enabled by recent advances in SRF technology. Like the original L2 modules, the HE modules will operate in continuous wave (CW) mode, meaning that achievement of high $Q_0$ in the cavities is crucial to keep the cryogenic heat load manageable. The minimum average $Q_0$ specification of $2.7 \times 10^{10}$ corresponds to a cryomodule dynamic 2 K heat load of 137 W at the nominal operating gradient of 20.8 MV/m.

*Table 1: Comparison of operating parameters for L2 (original superconducting linac) and HE (high energy upgrade) SRF cryomodules*

| Parameter | LCLS-II (L2) | LCLS-II-HE |
|---|---|---|
| $E_{acc}$ average | >16.0 MV/m | >20.8 MV/m |
| Total voltage | >133 MV | >173 MV |
| Dark current | <10 nA | <30 nA |
| Field emission x-rays | <50 mR/hr | <50 mR/hr |
| $Q_0$ nominal | >$2.7 \times 10^{10}$ | >$2.7 \times 10^{10}$ |
| Dynamic 2 K load | <80 W | <137 W |
| Peak detuning | <10 Hz | <10 Hz |
| Power coupler $Q_{ext}$ | $4.1 \times 10^7$ | $6.0 \times 10^7$ |

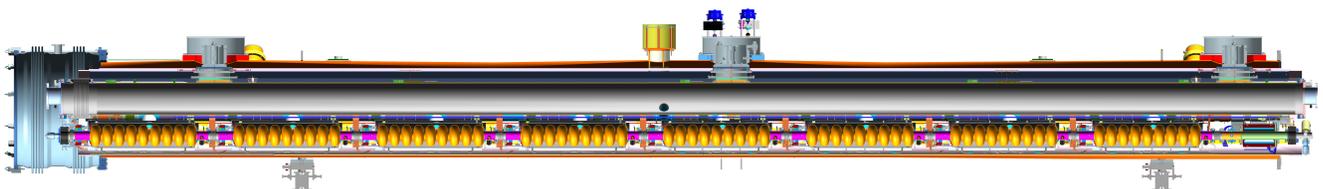

*Figure 1: Cross section of LCLS-II-HE cryomodule, which contains eight 1.3 GHz cavities. Each module is 12.2 m long.*

---

[a]Email: sposen@fnal.gov

To demonstrate these ambitious performance goals, a verification cryomodule (vCM) was assembled and tested at Fermilab using eight industrially produced nitrogen-doped cavities. The vCM is nearly identical in design to a production LCLS-II-HE cryomodule. The main difference is that the vCM is instrumented with additional temperature sensors on the tops and bottoms of Cavities 1, 4, 5, and 8, inside the helium vessels. The vCM was also intended to vet the production capabilities of the cavity vendor and the assembly and test procedures at Fermilab, as well as to evaluate operation at higher gradient and higher $Q_{ext}$, similar to how the prototype cryomodule (pCM) was used to qualify the L2 design [4]. In this paper, we discuss the performance of the vCM, including how some unique challenges to achieving higher gradients were addressed.

## II. CAVITY PROCESSING FOR HIGHER GRADIENTS

To achieve a 20.8 MV/m gradient on average in a cryomodule, the cavities need to first demonstrate at least this gradient in vertical test (VT). While many L2 cavities reached 21 MV/m in VT, Figure 2 shows that ~22% were limited below this gradient, and ~43% were limited below 23 MV/m (23 MV/m gives a ~10% margin on the gradient relative to the operating specification). This motivated a change to the cavity treatment process aimed at increasing the average sustainable gradient without sacrificing $Q_0$. An R&D program was initiated to develop a modified nitrogen doping process for this purpose. In place of the 2/6 process, two new process were developed and evaluated, 2/0 and 3/60 (2/6, for example, refers to the nitrogen injection in the vacuum furnace while the cavity is at 800 C, that is, 2 minutes of nitrogen exposure followed by 6 minutes of annealing before the furnace is turned off [5]).

The 2/0 process was eventually selected based on superior results from the R&D program, which also found that the modified doping process was more successful if the cavity received its post-furnace electropolishing (EP) step at a decreased temperature. The development of the processes for 2/0 doping and cold EP are reviewed elsewhere [6]–[8] and will be documented in a future publication. This program was crucial for achieving cavity performance that meets LCLS-II-HE requirements.

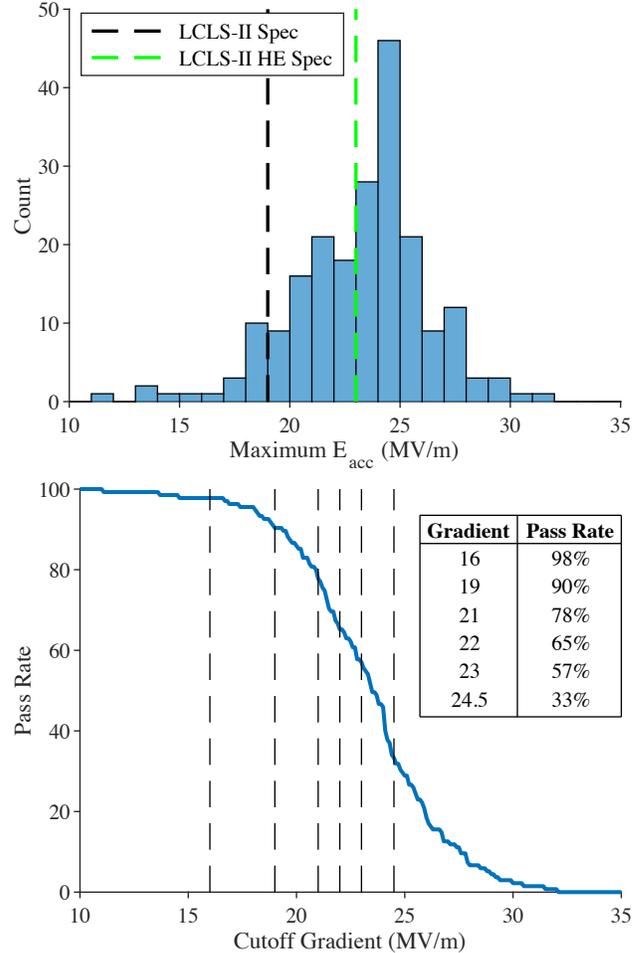

Figure 2: Statistics for vertical testing of production L2 cavities, showing that the gradient reach was not sufficient for HE. For L2, the minimally accepted gradient in VT was 19 MV/m; for HE it is 23 MV/m.

These treatment processes were industrialized via technology transfer as was done for the L2 cavities [9]. For the vCM, 10 cavities were produced and treated by the production cavity vendor using 2/0 doping and cold EP. Prior to doping, the cavities were given an elevated temperature heat treatment to improve flux expulsion, similar to what was done for the L2 cavities [10]. This is especially important for HE because sensitivity to trapped flux increases with increasing gradient [11]. All 10 cavities showed excellent performance, exceeding the minimum requirements, and 8 were selected for use in the vCM, as indicated in Figure 3.

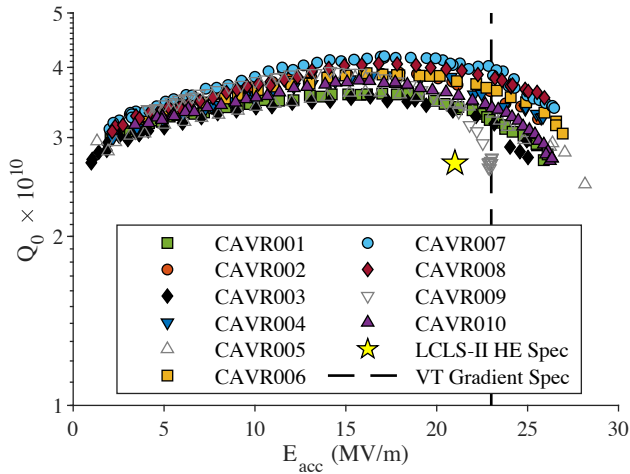

*Figure 3: Vertical test qualification of 10 cavities tested for the vCM. All 10 qualified and 8 were selected (open symbols indicate the two not selected).*

### III. L2 SPORADIC QUENCH EXPERIENCE

During tests of the L2 cryomodules, many cavities could reach gradients close to the administrative limit of 21 MV/m (as a precaution, cavities were not operated more than 5 MV/m above the nominal specification). In many cases, however, cavities would quench if left at gradients near 20 MV/m. Some cavities would remain stable for long durations at ~20 MV/m, but others would quench after several minutes or sometimes after several tens of minutes. Quenching is problematic in operation due to the disruption to the beam and from the $Q_0$ degradation that can occur (as explored in Section VII).

The 'sporadic quench' behavior led to a new requirement for the gradient to be considered 'usable.' In addition to requiring that any associated x-ray radiation be less than 50 mR/hr and that the cavity be 0.5 MV/m below the 'ultimate' quench limit, the cavity must also operate for more than one hour without quenching at the usable gradient. As a result, there were many cases in which the L2 cavities had usable gradients 1 MV/m or more below their maximum gradients, as shown in Figure 4. Cavities with field emission related radiation >50 mR/hr, poorly tuned higher order mode (HOM) notch filters, or similar extrinsic limitations were excluded for this comparison. Of those considered, 52 cavities (23%) have a usable gradient at least 1 MV/m lower than the maximum gradient, likely as a result of sporadic quenching. While this was not an issue for L2 as the average gradient still met the specification, it has the potential to be a limitation at the higher HE gradients.

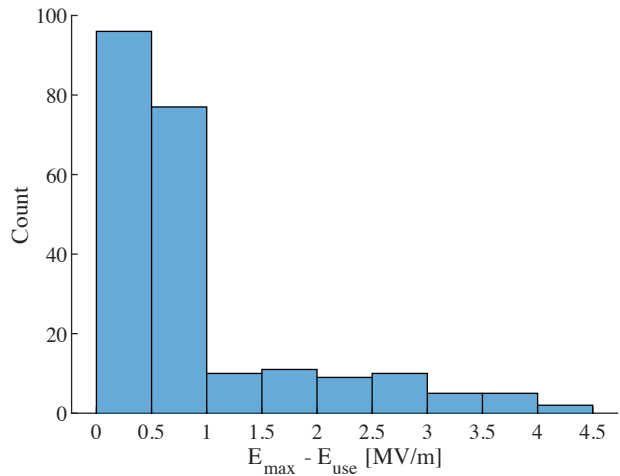

*Figure 4: Histogram of the maximum gradient minus the usable gradient for L2 cavities without excessive field emission, HOM port heating, or similar limiting extrinsic factors.*

Two mechanisms were investigated as possible causes of these quenches, namely end-group heating (e.g., from excess RF leakage through an HOM port due to its notch filter being poorly tuned) and multipacting. The investigation revealed no correlation between quench incidence and end-group temperature. However, there is supporting evidence for multipacting being the cause:
- Sporadic quenches were observed only in the multipacting band for TESLA shaped cavities, ~17-24 MV/m [12]
- Sporadic quenches typically coincided with a burst of x-rays, suggesting electron activity. Figure 5 illustrates this correlation for cavities from a Fermilab L2 module test
- Processing (repeated quenching) helps reduce the frequency of sporadic quenching. That is, the associated electron activity may reduce the secondary electron yield (SEY) of the niobium surface by removing adsorbed gases

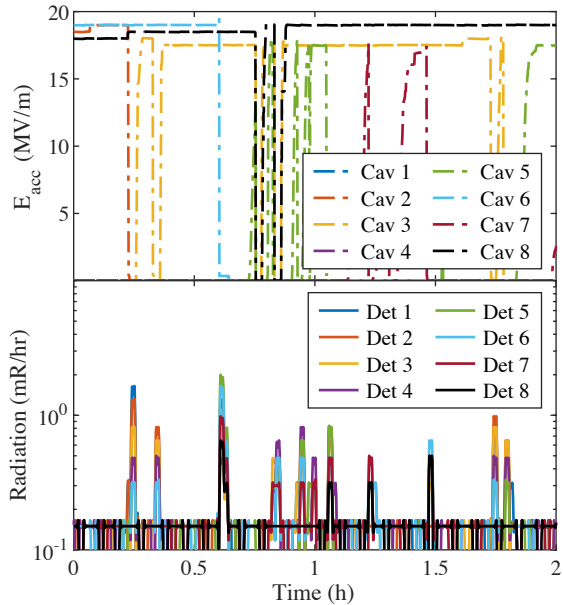

*Figure 5: A strong correlation was observed between quenching (appearing in the top plot as a sudden drop in gradient) and bursts of x-rays as observed with radiation detectors located alongside the cryomodule (bottom plot). The data are from Fermilab cryomodule 16 in the L2 production series.*

The fact that processing reduced the frequency of sporadic quenching was encouraging. To see if it could bring the usable gradient (i.e., stable for 1 hour) closer to the maximum gradient (i.e., stable for even a short duration), an experiment was carried out at the LERF facility at Jefferson Lab [13] with an L2 cryomodule. The cavity that was studied had an initial usable gradient of 17 MV/m, limited by sporadic quenching, but in VT it reached 24 MV/m. The gradient was then raised to the administrative limit of 21 MV/m and held there until a quench occurred. This process was then repeated approximately 15 times. The first quenches occurred after a few minutes while later quenches were spaced by up to tens of minutes. Afterward, the cavity operated stably at ~20 MV/m for more than 9 hours until it was turned off. The usable gradient was thus increased by at least 3 MV/m as a result of this quench processing [14].

This several-hour-long procedure was not implemented for the L2 cavities due to time constraints and the fact that the 16 MV/m specification was being met. As such, it was not clear if this technique could successfully process sporadic quenching in general nor whether it might return, for example, after more than 9 hours of operation. These questions would be investigated in the vCM.

In light of these observations, getter pumps were added to allow the vCM cavity string to be continuously pumped after assembly. The goal was to minimize residual gas condensation on the surface of the cavities during cooldown, which should reduce the SEY and therefore the likelihood of multipacting. When the vCM arrived at the test stand, the cavity string pressure was below $2 \times 10^{-8}$ Torr, multiple orders of magnitude lower than in previous Fermilab L2 cryomodules, which did not have continuous pumping after assembly.

## IV. INVESTIGATION OF SPORADIC QUENCHING IN VERTICAL TEST

To investigate the sporadic quench behavior and its relation to multipacting, vertical tests were performed at Fermilab on a single-cell 1.3 GHz cavity that was treated with 2/0 nitrogen doping and cold EP. It was tested twice, once with a temperature map (i.e., an array of sensors that measures the cavity surface temperature profile) and once with second-sound sensors (i.e., an array of oscillating super-leak transducers, which detect signals resulting from a cavity quench that travel through superfluid helium at a speed of about 2 cm/ms). Photographs from the two experiments are shown in Figure 6.

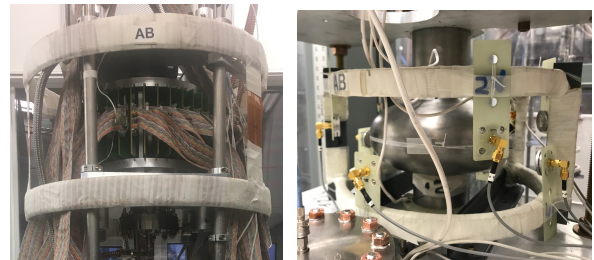

*Figure 6: Single cell cavity setup for evaluating sporadic quenching with a temperature map (left) and second sound sensors (right).*

As shown in Figure 7, the cavity was able to reach 25 MV/m without quenching and remain there for many minutes, but would sporadically quench if held at a gradient of 21 MV/m. The sporadic quenching near 21 MV/m occurred both before and after the stable period at 25 MV/m. The highest gradient achievable in the cavity, its 'ultimate quench field,' was 32 MV/m.

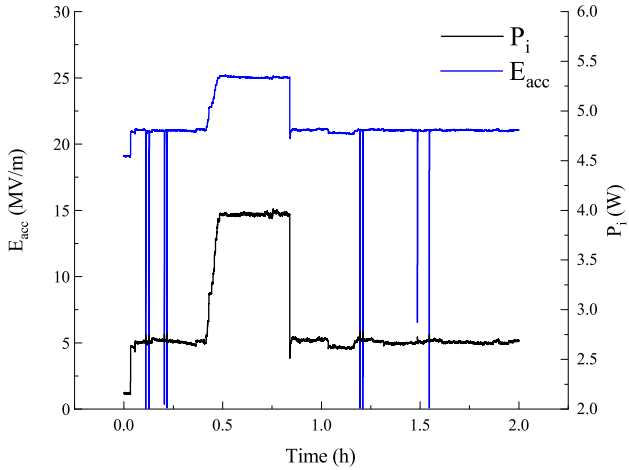

*Figure 7: Data taken during a vertical test of a single cell 1.3 GHz cavity. The rapid drops in gradient indicate when quenching occurred. $P_i$ is the cavity input RF power.*

Temperature map and second sound data for both the sporadic quenching and the ultimate quench behavior are shown in Figure 8. After sporadic quenches occurred, temperature maps showed changes in the heating in several areas near the equator (Thermometer #8), which is where electron impact from multipacting is expected to occur at this gradient [12]. The changes in heating suggest redistribution of trapped flux following quenching in these areas. The middle plot is an example of a temperature map measured in steady state at 21 MV/m after a sporadic quench. The bottom plot shows the external surface heating immediately following a quench at 32 MV/m (shortly enough afterwards that it measures the temperature increase from the quench itself), which appears to be the ultimate quench limit. The ultimate quench temperature profile is reproducible and shows highly localized heating compared to the more broadly distributed heating observed after a sporadic quench.

The second sound measurements corroborate this characteristic distinction between 'sporadic' and 'ultimate' quench behavior. When an ultimate quench occurs, the transducers closest to the quench spot show the earliest signal from the second sound, followed by those further away from the quench, consistent with a localized quench. But when a sporadic quench occurs, the second sound signals from all eight transducers arrive early, consistent with distributed quenching around the equator, rather than one location.

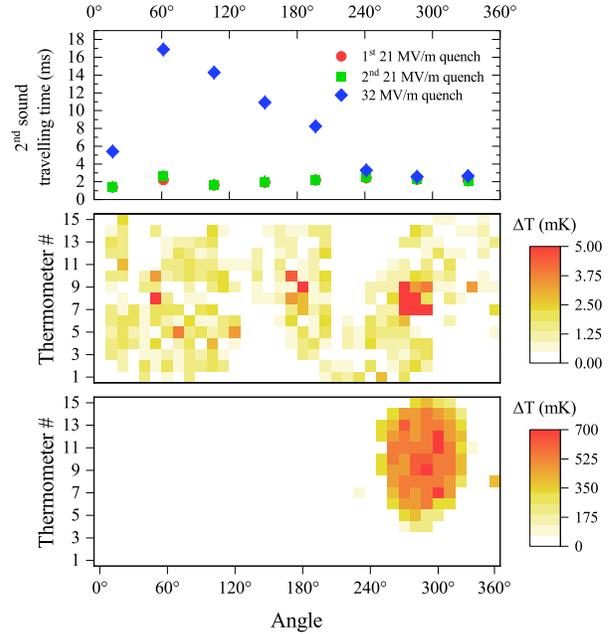

*Figure 8: Second sound (top) and temperature map data for the 1.3 GHz single cell cavity. The middle plot shows heating in steady state at 21 MV/m after a sporadic quench, and the bottom plot shows heating immediately following a quench at 32 MV/m.*

These measurements are consistent with similar ones performed on a $Nb_3Sn$ 1.3 GHz single cavity. It had showed quenching behavior at similar gradients where multipacting was the suspected cause. In this case, quench processing also led to the cavity reaching higher gradients [15]. The bilayer of $Nb_3Sn$ on Nb results in thermo-currents that trap flux, providing a clear indication in the T-map where the quench occurred. As shown in Figure 9, large heating occurred along the equator.

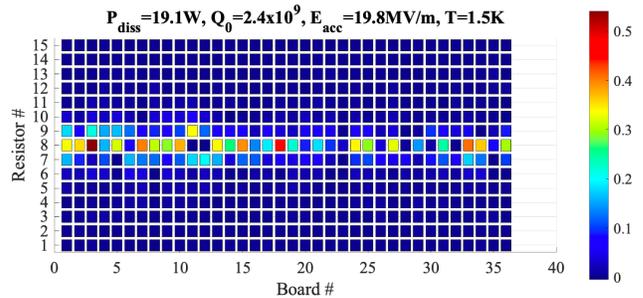

*Figure 9: Temperature map of a $Nb_3Sn$ single cell 1.3 GHz cavity after a sporadic quench. From [15].*

Multipacting in SRF cavities can manifest as a 'barrier' in gradient, where additional forward power results in the same gradient but lower $Q_0$, and the cavity does not quench unless excessive forward power is used. There are also observations of multipacting having weak barrier behavior but still causing quenching [16]. Ref. [17] notes that multipacting can be accompanied by x-rays if some of the low energy multipacting electrons escape the multipacting region and are accelerated by the high electric fields in the cavity.

The results presented here are mostly consistent with these observations, but it is not typical that quench and x-ray production are delayed up to tens of minutes after the start of operation. Some possible reasons for the delay include the time it takes for multipacting to be initiated (Ref. [18] discusses a possibly similar case), the time for electrons to escape from the multipacting region, and the time for layers of gas to accumulate on the RF surface (e.g., adsorbed gas released from the high power coupler).

## V. CRYOMODULE ASSEMBLY AND PREPARATION FOR TEST

Both the assembly and testing of the vCM took place at Fermilab. The assembly and preparation for the test was nearly identical to that for the L2 cryomodules [19] with a few key differences:
- Modification of some of the cleanroom string assembly procedures (e.g., attaching the cold coupler sections to the cavities during string assembly instead of beforehand)
- Addition of getter pumps (see Section III)
- Extra manifolds added to the module test stand for plasma processing carts
- 4 kW solid state amplifiers (SSAs) replaced with 7 kW SSAs

## VI. GRADIENT PERFORMANCE

The cryomodule was cooled down to 4 K over several days following procedures established for L2 cryomodule testing [20], then the helium return line was pumped down to achieve 2 K superfluid helium in the cavity helium vessels. After low power tests and setup (e.g., tuning cavities to 1.3 GHz), the cavity gradients were increased. It was quickly found that Cavity 1 showed unusual behavior in that the coupler vacuum pressure spiked at low gradients. This was later found to be caused by a loose connection between the fundamental power coupler warm and cold inner conductor assemblies. It was repaired in situ during a room temperature warmup partway through testing, after which, Cavity 1 behaved normally. This issue was not observed in the other 7 cavities, which each ramped up to 16 MV/m without issue.

After the initial gradient ramp to 16 MV/m, the cavities were individually brought to higher gradients, during which quenches occurred many times (typically several tens of quenches per cavity). The LCLS-II low level RF (LLRF) system [21] automatically disables the RF power when signals consistent with a quench are detected. Without this protection, the liquid helium would boil rapidly as the drop in $Q_0$ during a quench makes the RF coupler better matched to deliver power to the cavity. With the quench heat load greatly reduced, operators could quench a cavity many times within a short period, allowing for relatively fast processing, as shown in Figure 10.

Pulsed mode processing was used for Cavities 1 and 5 with a ~40 ms on time and a ~3 Hz repetition rate. During the periods that pulsing was applied (starting after the ~30 minute points in the plots), the gradient recorded by the logger was significantly lower than the peak gradient reached due to sample time limitations. For this reason, substantial radiation was observed during these periods at what appears to be a low gradient. Pulsing was also applied to Cavity 3 (at around 10 minutes into the plot) but quenching was not observed (nor radiation).

For Cavity 5, low level radiation was observed without corresponding cavity quenches, and the radiation increased rapidly with gradient, which is interpreted as a signature of field emission. For cavity 6, radiation was also observed without quenching, but it was present only briefly, not continuously. This could be an indication of some minor processing activity.

Some cavities exhibited stronger quenching behavior than others. For example, Cavity 1 showed frequent quenches after an hour of operation, and could not be driven above 20 MV/m in CW mode without quenches occurring. On the other hand, Cavity 4 reached higher than 20 MV/m without experiencing a single quench.

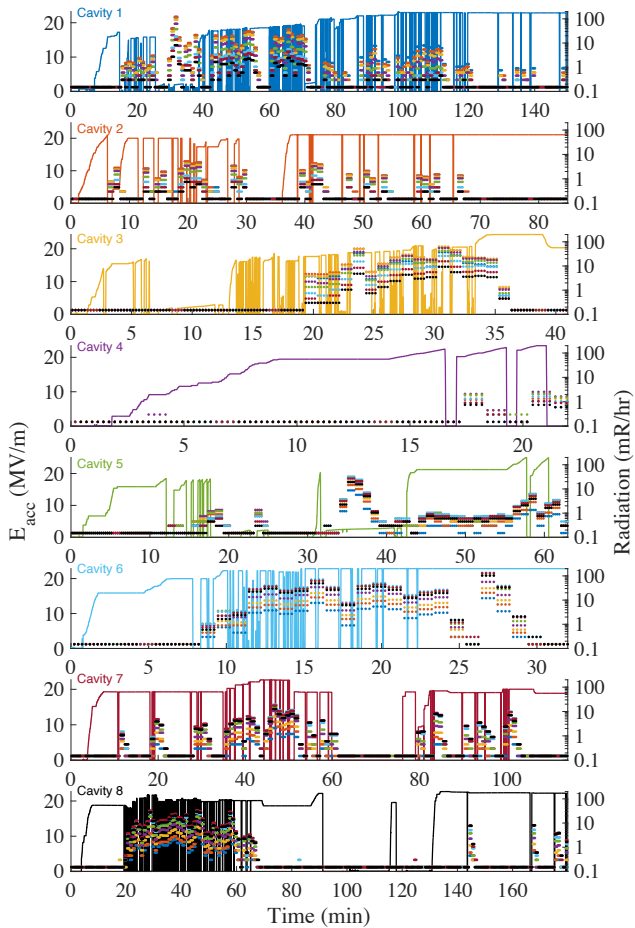

*Figure 10: Quench processing of the eight vCM cavities after the initial ramp to 16 MV/m. The solid lines are cavity gradients and the dots are readings from the radiation detectors located along the side of the module, which have an integration time of ~1 minute. Detector colors correspond to those in Figure 5. Some cavities have pauses in testing removed from the plots for ease of visualization.*

During initial processing, the goal was to achieve stable operation at ~20 MV/m for several minutes in each cavity, which took around 12 hours in total. Additional processing took much longer. For example, even if a cavity would operate stably for longer than 10 minutes at a time, several quenches might occur over the course of an hour. This behavior would typically persist until significant time was spent operating at an elevated gradient. An example of such processing is shown in Figure 11.

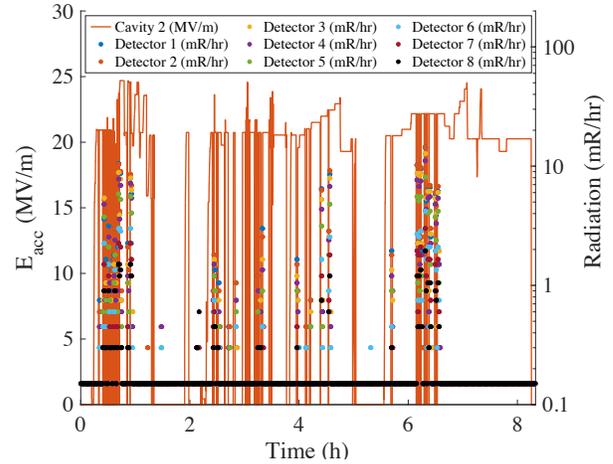

*Figure 11: Example of sporadic quenching over a long duration of operation at elevated gradient. A ~1hr pause in testing cavity 2 (during which other cavities were tested) is removed from the plot for ease of visualization.*

Figure 12 compares the maximum gradients measured in VT with the maximum and usable vCM gradients achieved by the end of the testing program. The maximum gradient during the cryomodule test was defined as the value at which a quench would consistently occur shortly after being attained, or the 26 MV/m administrative limit if that was reached without a quench (which was the case for Cavities 4 and 6). There is good agreement between the VT and vCM maximum gradients with a few percent difference for most cavities and a maximum difference of 9% for Cavity 1.

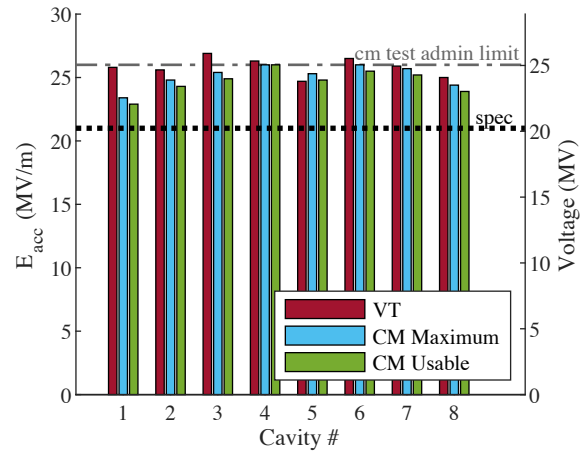

*Figure 12: VT maximum gradient, cryomodule maximum gradient, and cryomodule usable gradient for each cavity in the vCM. The uncertainty in the gradient measurements is approximately 10%.*

The usable gradient for each cavity was limited primarily by the requirement of a 0.5 MV/m buffer below ultimate quench. The usable gradient was measured one cavity at a time as there was a tendency for a sporadic quench in one cavity to result in quenches in multiple cavities. For the initial measurement of useable gradient, the gradient was lowered by 0.5 MV/m after a quench to reduce the likelihood of a quench occurring again. While this saved time, it resulted in usable gradients that were 2-3 MV/m below the maximum gradients. However, as described in Section VIII, an extended unit test was performed with all 8 cavities running simultaneously, which involved keeping the cavities operating at high gradients for many days. During this extended operation, there was a reduction of the incidence of sporadic quenches, and by the end of this unit test, cavities were able to operate within a few percent of their maximum gradients without sporadic quenches occurring. Several of the usable gradients shown in Figure 12 were measured at the end of the unit test when the cavities were pushed close to their maximum gradients.

Remarkably, little field emission was observed during the vCM test. Seven of the eight cavities produced no detectable radiation up to their maximum gradients. At 21 MV/m, Cavity 5 generated radiation of 0.6 mR/hr in a nearby detector, well below the maximum allowable limit. Moreover, the source of field emission in Cavity 5 was processed during later testing, after which the vCM was field emission free. This achievement is a credit to the established state-of-the-art cleanroom assembly practices and to the expertise of the cleanroom team. It is encouraging both for HE cryomodule production and for future high gradient SRF cryomodule production.

### VII. $Q_0$ MEASUREMENTS

The $Q_0$ of the cavities were individually determined at the nominal gradient (21 MV/m) based on a mass flow measurement of the helium boil-off gas. A heater was used for the mass flow calibration, and the static load was determined by measuring mass flow with the RF off. Before the $Q_0$ measurement, the vCM was thermally cycled, raising the temperature to ~50 K, then fast cooled with a ~75 g/s helium mass flow to achieve strong flux expulsion [22], [23]. The $Q_0$ results are shown in Figure 13 together with measurements from VT. The vCM $Q_0$ values are nearly all lower, possibly due to trapped flux, but the average is $3.0 \times 10^{10}$, well above the $2.7 \times 10^{10}$ spec. The uncertainty in the $Q_0$ measurements is approximately 15%.

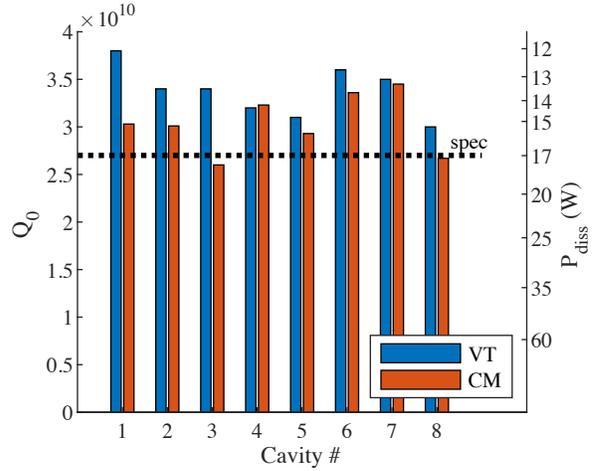

*Figure 13: $Q_0$ in the cryomodule compared to $Q_0$ in VT. Measurement were made at the nominal gradient (21 MV/m).*

Table 2 summarizes the performance of the vCM cavities including the $Q_0$ results and the gradient measurements discussed in Section VI.

*Table 2: Summary of cavity performance in vCM and VT. The VT $Q_0$ values are corrected for stainless steel flange losses (0.8 nΩ subtracted, based on simulations). 'FE onset' refers to the lowest gradient at which field emission was observed at the end of the run.*

| | | Vertical Test | | Cryomodule Test | | | |
|---|---|---|---|---|---|---|---|
| Slot in CM | Serial # CAVR | $E_{max}$ (MV/m) | $Q_0/10^{10}$ at 21 MV/m | $E_{max}$ (MV/m) | $E_{use}$ (MV/m) | FE onset (MV/m) | $Q_0/10^{10}$ at 21 MV/m |
| 1 | 004 | 25.8 | 4.3 | 23.4 | 22.9 | none | 3.0 |
| 2 | 002 | 25.6 | 3.7 | 24.8 | 24.3 | none | 3.0 |
| 3 | 006 | 26.9 | 3.8 | 25.4 | 24.9 | none | 2.6 |
| 4 | 010 | 26.3 | 3.6 | 26.0 | 26.0 | none | 3.2 |
| 5 | 001 | 24.7 | 3.4 | 25.3 | 24.8 | none | 2.9 |
| 6 | 007 | 26.5 | 4.0 | 26.0 | 25.5 | none | 3.4 |
| 7 | 008 | 25.9 | 3.9 | 25.7 | 25.2 | none | 3.4 |
| 8 | 003 | 25.0 | 3.3 | 24.4 | 23.9 | none | 2.7 |
| Avg | | 25.8 | 3.8 | 25.1 | 24.8 | | 3.0 |

Prior to performing the 75 g/s cooldown, the cavities were allowed to quench numerous times while operating for hours at gradients above 20 MV/m. The goal was to decrease the likelihood that they would quench after cooldown when the $Q_0$ measurements were made since quenching can trap ambient flux and reduce $Q_0$ [24]. Measuring $Q_0$ requires the cavities to operate at 21 MV/m for tens of minutes so that the helium mass flow stabilizes. Despite the pre-conditioning, three cavities quenched during the measurements. Fortunately, each cavity operated stably long enough for the mass flow to stabilize, and by measuring the mass flow between quenches, multiple $Q_0$ measurements were made. Figure 14 shows an example of the data taken and Figure 15 summaries the $Q_0$ evolution with quench number for three cavities. Although this was not meant to be a systematic study of $Q_0$ degradation, it does offer insight into its size for these particular cavities. In this case, the corresponding surface resistances are seen to increase from 0.2 n$\Omega$ to 0.7 n$\Omega$ per quench, likely from flux trapping.

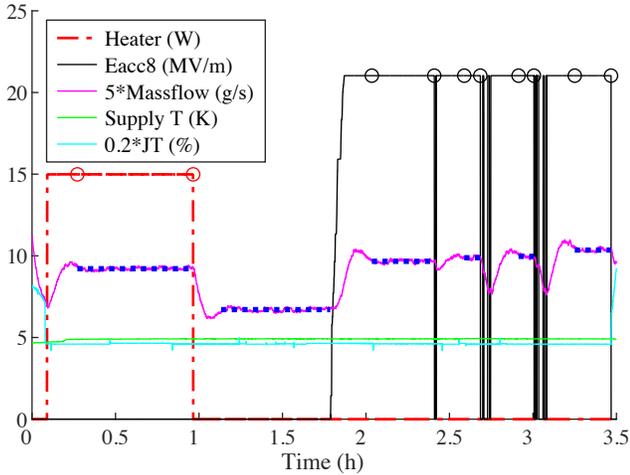

Figure 14: Data recorded during the $Q_0$ measurement of Cavity 8. Multiple quenches occurred during the process, but there were stable intervals long enough that allowed multiple $Q_0$ measurements to be made. This plot also shows the heater and static load calibration steps. The regions where the mass flow was evaluated are highlighted with circles and dotted blue lines. The helium inlet conditions (Joule Thomson valve setting and supply temperature) were held as constant as possible during this period.

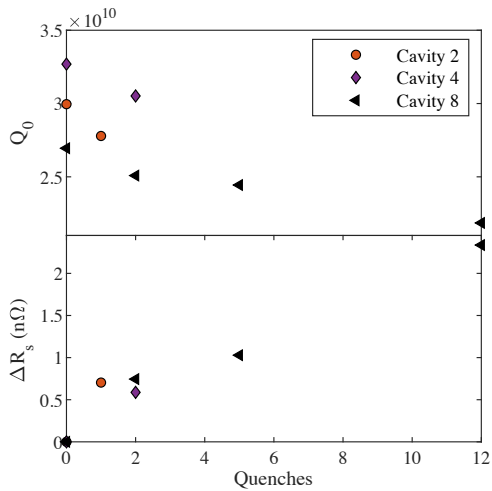

Figure 15: $Q_0$ versus quench number for three cavities (top) and the corresponding change in surface resistance relative to zero quenches (bottom). The degradation was recovered after thermal cycling.

During the unit test (see Section VIII), all eight cavities were operated simultaneously with the goal of maintaining an average gradient of 20.8 MV/m. Quenches occasionally occurred early in the test, but after a few days, no further quenches were observed. The number of quenches observed since the most recent thermal cycle was recorded for each cavity, and the $Q_0$ of each cavity was measured at the end of the unit test. These $Q_0$ results are compared to the earlier data in Figure 16 as a function of the number of quenches. It should be noted that these measurements were performed after different thermal cycles, which may have had different amounts of flux trapped during cooldown. The uncertainty in the $Q_0$ measurement was higher than usual, likely due to cryogenic instabilities in the helium supply system. The average $Q_0$ after the unit test was $(2.7 \pm 0.3) \times 10^{10}$.

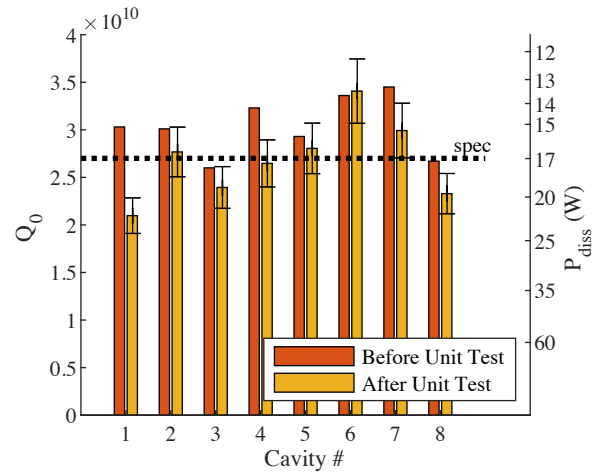

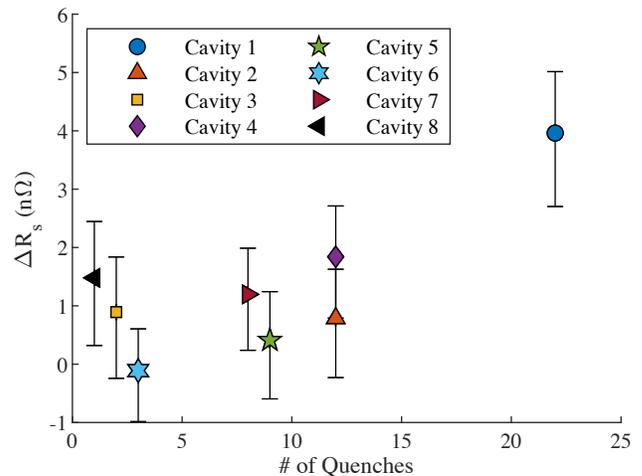

Figure 16: $Q_0$ comparison before and after the unit test (top) and the corresponding changes in surface resistance versus the number of quenches that occurred since the most recent thermal cycle (bottom).

These $Q_0$ degradation measurements show the importance of processing sporadic quenches as fully as possible and performing a final thermal cycle to recover $Q_0$ prior to long-term operation.

The effectiveness of the thermal cycle in expelling flux depends on the temperature gradient achieved across the cavity at the critical temperature, which is a function of the helium flow rate during cooldown. The average $Q_0$ for Cavities 2-8 after a 32 g/s cooldown (when Cavity 1 was non-operational) was just 1% lower than that after a later 75 g/s cooldown, indicating that the planned 32 g/s fast cooldown rate in the SLAC linac should be adequate. Figure 17 shows that the temperature difference from the top to the bottom of the cavities (which drives flux expulsion) is similar (or even higher) in the vCM compared to that in the pCM for the same mass flow. These measurements use temperature sensors inside the helium vessels, which were added to the vCM, but will not be in production cryomodules.

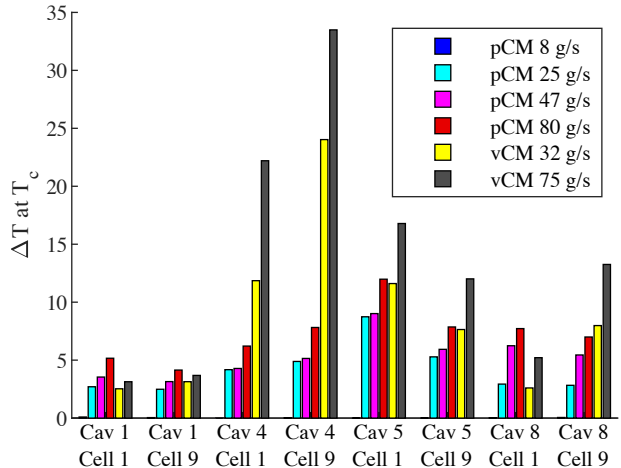

*Figure 17: Top to bottom cavity temperature difference at the critical temperature during cooldown. Having a high ΔT is important to achieve strong flux expulsion.*

## VIII.  UNIT TEST

A 12 day 'unit test' was carried out to evaluate the vCM performance with all eight cavities operating at a sufficiently high average gradient (20.8 MV/m) to achieve the HE cryomodule voltage specification of 173 MV. During this period, the LLRF system regulated the cavity amplitudes and phases and adjusted the piezo tuner voltages and RF input power to compensate for slow and fast cavity detuning variations. This operation mode is referred to as SELAP (Self-Excited Loop – Amplitude & Phase). Two issues of note were encountered early into the test, which are discussed below.

The liquid helium level in the cryomodule reservoirs was extremely stable when powering cavities individually or a few at a time, but when powering all 8 cavities simultaneously, it became difficult to maintain stability. The liquid is supplied to the cryomodule via a Joule-Thomson (JT) valve, which injects a gas and liquid mixture into the two-phase pipe that spans the length of the cryomodule. Upstream of the JT valve, the temperature of the high pressure helium was higher (5 K) than optimal (~ 3 K), which caused a significant amount of helium vapor ('flash') to be generated upon injection. This created a relatively large helium gas load, which propagated through the two-phase pipe in addition to the gas generated by the heating of the cavities.

Complicating matters, the CM was tilted by 0.5% on the test stand (6 cm over 12 m) to mimic the slope of the LCLS-II tunnel at SLAC. Thus, the liquid helium levels in the reservoirs at each end of the 2-phase line were not equal. With all 8 cavities operating, a 'vapor damming' effect occurred on occasion, making it difficult to supply sufficient liquid to all cavities. This appeared as a slow decrease in the upstream (higher elevation) reservoir level relative to that in the downstream one. The upstream level would not recover until the heat load of Cavity 1 (upstream most cavity) was lowered significantly. Figure 18 shows an example in which the Cavity 1 and 2 gradients were both lowered. To avoid such occurrences, Cavity 1 was operated at 16 MV/m for the unit test and the gradients of the other seven cavities were increased to 21.7 MV/m to meet the 173 MV target voltage. In the SLAC linac, the helium supply temperature is expected to be closer to 3 K, which should reduce the vapor generated and hopefully eliminate the need to adjust the gradients.

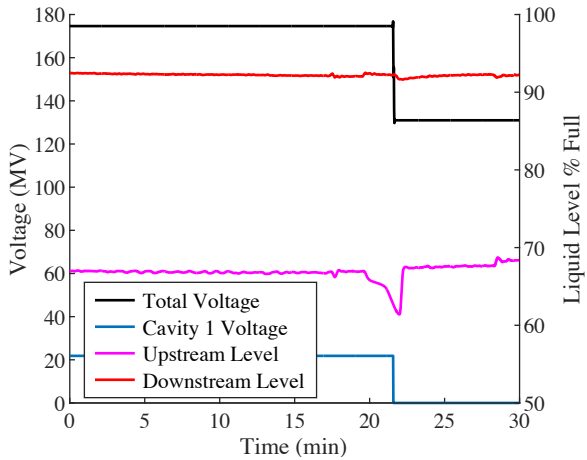

*Figure 18: Data showing that lowering the gradient in Cavities 1 and 2 recovered the liquid level drop in the upstream reservoir.*

Another issue encountered during the unit test involved the vacuum pressure level in the 'warm' coupler sections. These sections are pumped from a common manifold that runs along the outside of the CM vacuum vessel. When the cavities were operated in frequency tracking mode (SEL), the input RF power stayed fairly constant and the manifold pressure stabilized in the $10^{-9}$ Torr range. However, when the cavities were operated in SELAP mode, spikes in coupler vacuum pressure would occur as the RF input power increased by various amounts to compensate detuning. These outgassing bursts were likely due to the higher power and the shifts in the standing-wave pattern along the couplers, which exposed unprocessed regions to higher fields (nearly all RF power reflects from the cavities and the relative phase of the forward and reflected RF varies with the cavity detuning). The spikes were only observed for some cavities. Cavity 1 had a substantially higher level of spikes than the other cavities, which was not unexpected as it had been vented to fix its warm-to-cold inner conductor connection.

The vacuum spikes would sometimes exceed the $5\times 10^{-7}$ Torr fault level, shutting off RF power to all cavities – examples are shown in Figure 19. As a mitigation, the LLRF system reactive power overhead limit was set relatively low, ~15% of full scale, so large cavity detuning would cause the LLRF system to switch to frequency tracking mode, which reduced the RF power while maintaining the gradient. With this change, the coupler vacuum spikes were less frequent although the lowering of the Cavity 1 gradient to prevent liquid level drops also helped. However, the brief transitions to SEL mode caused cavity phase and amplitude deviations that exceeded LCLS-II specifications.

To reduce the RF-related outgassing, each of the couplers was RF processed in situ after the unit test. To allow higher power operation without quenching a cavity, the coupler external quality factor ($Q_{ext}$) was reduced, which lowered the gradient for a given input power. As discussed in Section XI, this processing allowed the reactive power limits to be increased.

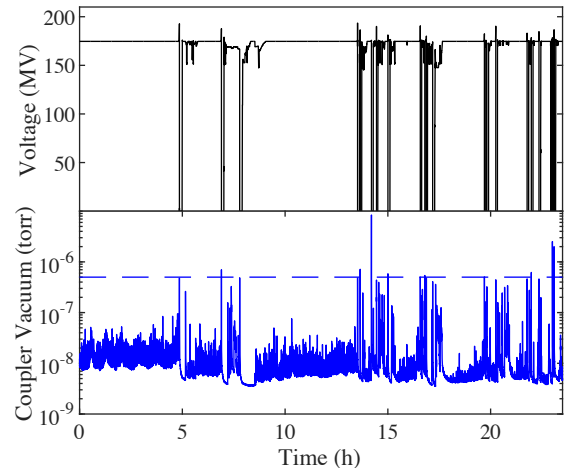

*Figure 19: Cryomodule voltage (top) and coupler vacuum level (bottom) verses time. Vacuum spikes that exceed the trip level (dashed line) shut off RF power to all cavities.*

Figure 20 is a plot of the total cryomodule voltage, upstream liquid level, and coupler vacuum pressure during a four day period of the unit test. The goal was to minimize interruptions while the cavities operated in SELAP mode at a combined voltage of 173 MV. An 84% uptime was achieved over the period plotted in Figure 20, where the majority of interruptions were due to upstream liquid level drops. Other down periods were due to a coupler vacuum trip, a brief conflict for cryogenic resources with another cryomodule being tested, and intentional pauses for adjustments, e.g., increasing the coupler $Q_{ext}$ values to offset the decreases that occurred from the gradual coupler temperature rise over time.

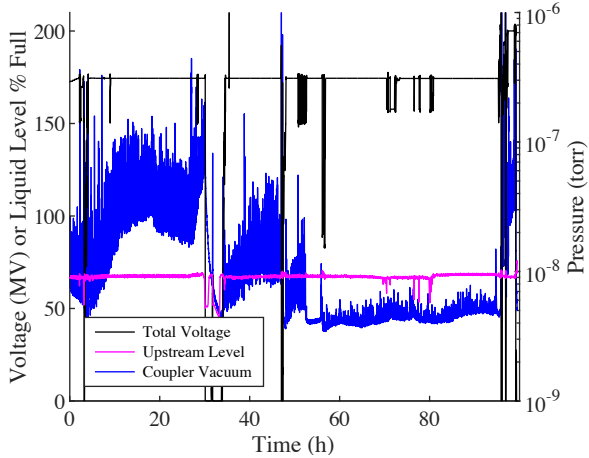

*Figure 20: Cryomodule voltage, upstream liquid level and coupler pressure over four days during the unit test. The decrease in the coupler vacuum pressure over time was due in part to $Q_{ext}$ adjustments that lowered the RF input power.*

On the last day of the unit test (and immediately following the period in the Figure 20 plot), the cavity gradients were increased to achieve a 200 MV CM voltage with the cavities operating in SELAP mode. Figure 21 shows the CM voltage and coupler vacuum pressure during this study. An even higher voltage may have been possible if not for the liquid level issue. During the ramp up to higher voltage, the Cavity 5 field emission site was processed, which was accompanied by a spike in the cavity vacuum pressure.

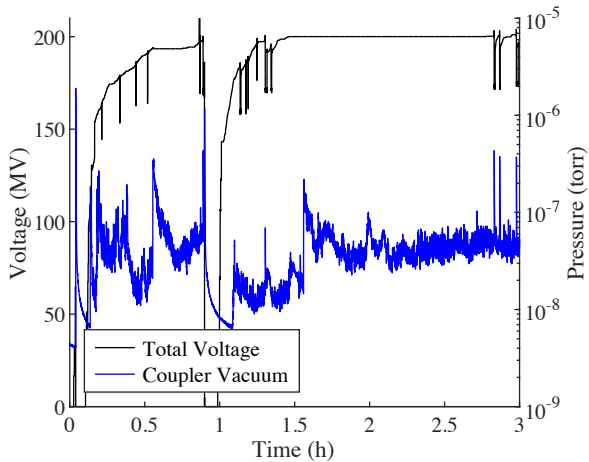

*Figure 21: Data taken during the high gradient test in which a total voltage of 200 MV was maintained for more than one hour.*

## IX. MICROPHONICS

During the unit test, the cavity frequency detuning was recorded at a 1 kHz sample rate for several hours at a time. The top plot Figure 22 is a histogram of detuning measurements during a typical scan, which in this case lasted 4.5 hours. The HE goal is for the rms detuning to be less than 1.7 Hz, which corresponds to 10 Hz at six sigma. At this rms limit, only one cavity per day is expected to exceed 10 Hz detuning in the HE linac. With the 7 kW SSAs that will be used to power the HE cavities, a gradient of 26 MV/m can be achieved with 10 Hz detuning while accelerating the nominal beam current of 30 µA. At lower gradients and beam currents, larger detuning can be accommodated.

During the tests of the L2 cryomodules at Fermilab, the average rms detuning was about 1.7 Hz. During the vCM test, the cryogenic system appeared to be less stable and the rms detuning of all cavities exceeded 1.7 Hz. The integrated rms detuning curves shown in the bottom plot of Figure 22 indicate that the largest contribution was from ~20 Hz vibrations that coupled to all the cavities. If these 20 Hz components were removed, the rms detuning values would be significantly reduced.

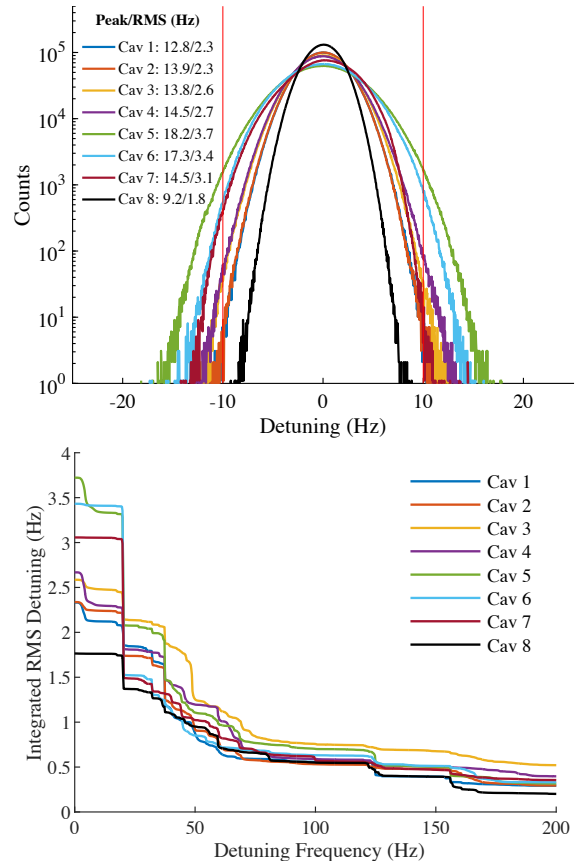

*Figure 22: (Top) Histogram of the detuning of the eight cavities over a 4.5 hour period. The goal is not to exceed ±10 Hz (red lines) at 6 sigma. (Bottom) Integrated rms detuning for the eight cavities from the frequency computed to 500 Hz.*

The 20 Hz vibrations were likely caused by thermal acoustic oscillations (TAOs) in the supply line. As shown in Figure 23, the frequency variations closely followed the 2.2 K helium supply line pressure changes. This is consistent with cavities 5, 6, and 7 having the largest 20 Hz detuning contribution as they are closest to the JT valve. The 20 Hz vibrations are believed to be due to an atypical instability in the local cryogenic system, and are not expected to occur in the HE linac. This instability will be investigated during the next cooldown.

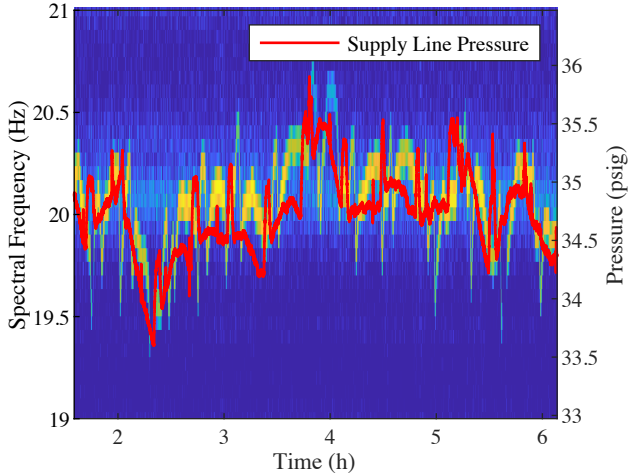

*Figure 23: Spectrogram of the Cavity 6 detuning near 20 Hz and the variation in the 2.2 K helium supply line pressure.*

Other contributions to the cavity detuning were from vacuum pumps that generated 15 Hz and 30 Hz vibrations. These sources were found by comparing the detuning spectra with the pumps on and off. Precautions have been taken to avoid this issue in the HE linac.

During the unit test, the JT valve was manually operated and very few adjustments (a few per day) were required to maintain stability. As a result, vibrations from JT valve motion were not a significant source of microphonics. Automatic control of the JT valve requires tuning of the feedback algorithm to prevent oscillations, which has not yet been done.

## X. OFF FREQUENCY OPERATION

The HE project requires that the cavities be able to operate up to 465 kHz below the nominal 1.3 GHz frequency as means of achieving multiple beam energies concurrently if the need arises [25]. The range of the cavity mechanical tuners were increased for this purpose and multi-cycle testing was done to qualify their lifetime [26]. As an additional validation test, the frequency of each cavity in the vCM was lowered by 465 kHz and they were then operated in SELAP mode for 3 hours.

No major issues were encountered although the fundamental mode RF leakage from the higher order mode (HOM) coupler ports slightly exceeded the specification in some cases. This can be avoided by lowering the target frequency somewhat when the ~1 MHz bandwidth HOM notch filters are tuned during HE cryomodule production.

## XI. REACTIVE POWER LIMIT

Prior to the off frequency tests, the couplers had been processed at low $Q_{ext}$ as discussed above. To see if the coupler vacuum spikes had been sufficiently suppressed to allow SELAP operation at larger cavity detuning, the reactive power limit was slowly increased for each cavity during its 3 hour off-frequency test. If the limit was exceeded at any time during the ~ 0.3 sec long waveforms that are continuously recorded by the LLRF system, a fault indicator was set, but the cavities continued to operate. The resulting fault rate vs reactive power limit is plotted in Figure 24 for several cavities.

In some cases, the fault rate appears to increase with reactive power, which seems counterintuitive. However, when the reactive power limit is low, the cavities can exceed the limit for extended periods, resulting in fewer recorded fault transitions. Using a 30% fractional reactive power limit largely eliminated the faults, and importantly, the processed couplers operated stably without causing vacuum trips.

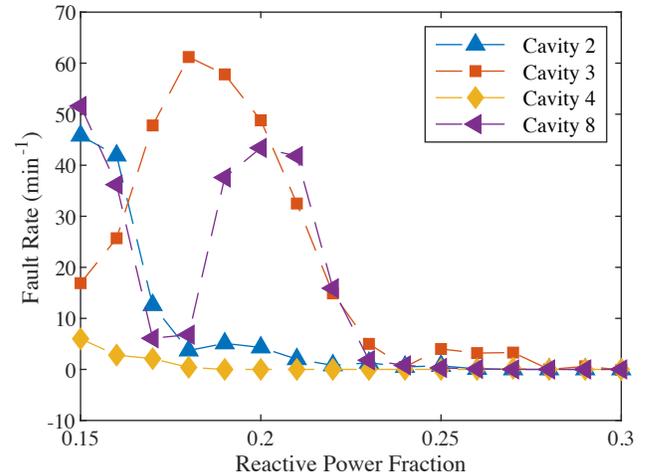

*Figure 24: Reactive power fault rate versus the reactive power limit for four cavities after coupler processing at low $Q_{ext}$.*

## XII. PLASMA PROCESSING

After the off-frequency test, the vCM was warmed to room temperature and plasma processing procedures were initiated (the procedures are described in Refs. [27]–[29]). After plasma processing, the vCM will be cooled down again to evaluate any changes in performance. Given the final emission-free state of the vCM, the effectiveness of plasma processing in reducing field emission cannot be evaluated. However, if the performance is not degraded, plasma processing will likely be tried again if field emission becomes an issue during cryomodule testing or possibly used in a retrofit process to modules from the SLAC SRF linac if they show substantial field emission.

## XIII. SUMMARY

The LCLS-II-HE verification cryomodule performance exceeded specifications. A CW voltage of 200 MV was achieved with the eight cavities operating simultaneously in SELAP mode for more than one hour. This voltage is higher than that achieved in any previous cryomodule operating CW. The cavity usable gradients averaged 24.8 MV/m and ranged from 23.0 MV/m to 26.0 MV/m, the latter being the administrative limit.

When testing LCLS-II cryomodules, the usable gradient was often limited by 'sporadic quenches,' that is, quenches that occurred suddenly under otherwise stable conditions. This phenomenon was also observed in the vCM cavities, but mitigated by extended processing during which repeated quenching occurred. As a result, higher usable gradients could be reached. Based on evidence from both cryomodule and single cell cavity tests, sporadic quenching appears to be due to equatorial multipacting, which is not unexpected at the gradients required for HE.

The average cavity $Q_0$ at 21 MV/m was $3.0 \times 10^{10}$, validating that the 2/0 nitrogen doping process maintains a high $Q_0$ in addition to improving the gradient performance. Measurements of $Q_0$ degradation after quenching emphasized the importance of mitigating sporadic quench behavior. Little field emission was observed during the tests, and this was processed during extended operation, eliminating all detectable radiation and dark current up to the maximum gradient of each cavity. This level of field emission control is promising not just for HE cryomodule production, but also for other SRF projects with challenging gradient specifications, such as ILC. The overall high gradient and high $Q_0$ performance is a promising milestone for HE and future high duty factor SRF linac applications.

## XIV. ACKNOWLEDGEMENTS

The authors would like to thank the many people who contributed to the results presented in this paper. This includes the teams responsible for designing, assembling, and testing the cryomodules at Fermilab, the Fermilab cavity assembly and vertical test teams, and the LCLS-II project teams at Fermilab, Jefferson Lab, LBNL, and SLAC. Thanks to Mike Drury and the Jefferson Lab cryomodule testing team for hosting Fermilab and SLAC visitors during the cryomodule measurements at LERF that are discussed in Section III. Thanks also to Hasan Padamsee and Sergey Belomestnykh for useful discussions about multipacting. This work was supported by the United States Department of Energy, Offices of High Energy Physics and Basic Energy Sciences under Contracts DE-AC05-06OR23177 (Fermilab), DE-AC02-76F00515 (SLAC), and DE-AC02-05CH11231 (LBNL).